\documentclass{article}

\usepackage{filecontents}
\usepackage{amsmath}
\usepackage{amssymb}
\usepackage{authblk}
\usepackage[backend = biber, bibencoding = utf-8, sorting = none]{biblatex}
\usepackage[margin=1in]{geometry}
\usepackage{enumerate}
\usepackage{graphicx}
\usepackage{setspace}
\usepackage{url}
\usepackage{xr}
\newcommand{\suppref}[1]{Supplementary Material Section~\ref{Supp-#1}}
\bibliography{./references.bib}
\title{Change Point Detection of Events in Molecular Simulations using \pkg{dupin}}
\date{\today}

\author[a]{Brandon L. Butler}
\author[a]{Domagoj Fijan}
\author[a,b]{Sharon C. Glotzer\textsuperscript{*}}

\affil[a]{Department of Chemical Engineering, University of Michigan, Ann Arbor, MI.}
\affil[b]{Biointerfaces Institute, University of Michigan, Ann Arbor, MI, USA}
\affil[*] {Corresponding author\\ \textit{E-mail address:} sglotzer@umich.edu}

\DeclareMathOperator*{\argmin}{\arg\,min}
\newcommand{\pkg}[1]{\texttt{#1}}

\begin{document}
    \onehalfspacing
    \maketitle

    \begin{abstract}
        Particle tracking is commonly used to study time-dependent behavior in many different types of physical and chemical systems involving constituents that span many length scales, including atoms, molecules, nanoparticles, granular particles, and even larger objects.
Behaviors of interest studied using particle tracking information include disorder-order transitions, thermodynamic phase transitions, structural transitions, protein folding, crystallization, gelation, swarming, avalanches and fracture.
A common challenge in studies of these systems involves change detection.
Change point detection discerns when a temporal signal undergoes a change in distribution.
These changes can be local or global, instantaneous or prolonged, obvious or subtle.
Moreover, system-wide changes marking an interesting physical or chemical phenomenon (e.g.\ crystallization of a liquid) are often preceded by events (e.g.\ pre-nucleation clusters) that are localized and can occur anywhere at anytime in the system.
For these reasons, detecting events in particle trajectories generated by molecular simulation is challenging and typically accomplished via \textit{ad hoc} solutions unique to the behavior and system under study.
Consequently, methods for event detection lack generality, and those used in one field are not easily used by scientists in other fields.
Here we present a new Python-based tool, \pkg{dupin}, that allows for universal event detection from particle trajectory data irrespective of the system details.
\pkg{dupin} works by creating a signal representing the simulation and partitioning the signal based on events (changes within the trajectory).
This approach allows for studies where manual annotating of event boundaries would require a prohibitive amount of time.
Furthermore, \pkg{dupin} can serve as a tool in automated and reproducible workflows.
We demonstrate the application of \pkg{dupin} using two examples and discuss its applicability to a wider class of problems.

    \end{abstract}

    \section{Introduction}%
\label{sec:intro}
Computer simulations of molecular systems from the atomic to colloidal particle scale are a cornerstone of modern materials research.
Given ongoing improvements in algorithms and processor speed, newer studies may make routine use of thousands of simulations or more~\cite{thomas.etal2018,thompson.etal2019}.
As a result, much work has been done to automate, streamline or otherwise simplify the management of large-scale simulation studies from software that manages data and workflows~\cite{adorf.etal2018b,huber.etal2020} to packages used in data pipelines~\cite{ramasubramani.etal2020,dice.etal2019,mcgibbon.2015,agrawal.2011,gowers.2016,stukowski2009}.

We continue this trend by developing a software package that detects events (transitions) within point cloud data, the kind of data produced in molecular simulations.
Event detection in point cloud data from particle trajectories is difficult because \textit{a priori} information on the nature of the transition, identifying features, or precursors of the transition (if any) is often missing.
For studies involving multiple transitions or pathways, as well as large systematic studies, detection is further complicated by the need to automate the task.
This paper addresses this problem using approaches collectively known as change point detection (CPD)~\cite{aminikhanghahi.cook2017}.

The techniques of CPD, which are commonly used in signal processing, have yet to infiltrate materials research~\cite{aminikhanghahi.cook2017}.
Change points are defined by abrupt changes beyond expected fluctuations in a time-resolved dataset.
In this paper, we will use particle trajectory data as our starting dataset.
Molecular dynamics simulation as well as experiments using dynamic (e.g.\ liquid phase) transmission electron microscopy~\cite{luo.etal2017,ou.etal2020} or confocal laser scanning microscopy~\cite{royall.etal2007,abdel-hafez.etal2018} --- in conjunction with particle tracking software --- can produce data consisting of time resolved particle positions and orientations.

Many algorithms~\cite{aminikhanghahi.cook2017} have been introduced to detect events associated with change points in data for the purpose of monitoring human activity~\cite{aminikhanghahi.cook2017b}, determining useful telemetry in data centers~\cite{alves.etal2020} or predicting machine degradation~\cite{shi.chehade2021}.
Change point detection approaches can be categorized as supervised~\cite{feuz.etal2015,han.etal2012,reddy.etal2010} or unsupervised~\cite{aue.etal2009,barry.hartigan1993,kawahara.masashi2012,kawahara.etal2007,keogh.etal2001} as well as offline or online.
Supervised methods require external labelling of events to \textit{train} on, while unsupervised do not.
Offline methods require the entire dataset as input.
In contrast, online methods analyze the data as a stream while data is generated so that signals are screened in real time.
The approach we take casts event detection as an optimization problem where the objective is minimized over change point locations~\cite{bosc.etal2003}.

Current approaches for detection of interesting events in molecular simulations involve the use of system or particle level order parameters.
After computing these features, order parameters or environment labels, events are detected using system-specific or problem-specific detection schemes or by visual inspection.
Automating the detection of meaningful events would significantly improve current workflows, facilitate ``big data'' studies in materials research, and provide a consistent approach across studies.

The first step for the approach we take is to generate a set of descriptors (usually order parameters) which capture structural features of the system that might undergo a change during the simulation.
Steinhardt order parameters\cite{steinhardt.etal1983} (including its derivatives such as Minkowski structure metrics (MSM)~\cite{mickel.etal2013}) are an extremely versatile type of order parameters used to describe local structure of matter based on an expansion of bond orientational order in the basis of spherical harmonics.
They are used in all fields of molecular simulations, spanning simulations with both atomistic and coarse grained potentials to study liquids,\cite{palmer.etal2022} crystals,\cite{steinhardt.etal1983,lechner.dellago2008} and other phases.
Depending on types of phases studied other order parameters such as nematic order parameter, local density can also be employed.
In the atomistic simulations Smooth Overlap of Atomic Position (SOAP) \cite{bartok.etal2013} is another commonly used order parameter, including recent additions of time-dependent SOAP variants\cite{Caruso.etal2023}.
These local order parameters can be used to classify particles into environments using machine learning (ML)~\cite{dice2021,boattini.etal2019,schoenholz.etal2016,adorf.etal2020,dietz.etal2017,spellings.glotzer2018,barnard.etal2023}.
Current practices in molecular simulations for event detection involve system or particle level order parameters.

In this paper, we present a new, open-source Python package \pkg{dupin} (named after Edgar Allen Poe's detective C. Auguste Dupin) for generic, autonomous event detection in particle trajectories with local or system-wide transitions.
We show how \pkg{dupin} can partition a system's trajectory into regions of transition and stable (or metastable) states through the use of generic order parameters.
In the following section, we outline \pkg{dupin}'s multi-stage procedure for detecting a set of change points from a system trajectory.
We discuss options available to the user at each stage of the process.
We then present two example applications using \pkg{dupin} and show its utility in determining the temporal bounds of system-wide structural transitions and particle-level events.
We conclude with a general discussion of \pkg{dupin} and potential extensions.
\pkg{dupin} is available on GitHub and distributed through conda-forge and the Python Package Index (PyPI).

\section{Results}%
Primary purpose of \pkg{dupin} is detection of events in time correlated point cloud data sets (particle trajectories).
We propose and implement an approach in which we translate a set of point cloud data into a set of relevant descriptors representative of each frame of the trajectory which we refer to as the signal.
These descriptors should be carefully selected to capture critical aspects of the system's dynamics, to ensure successful CPD process.
Once such a signal is generated, we can apply change point detection (CPD) algorithms to detect events in the trajectory.
To do this we have developed a multi-step scheme which prepares the raw point cloud input into a form suitable for event detection.
\pkg{dupin} introduces novel methodologies for some steps of this workflow while also supporting a broad range of established methods and user-defined functions.
This flexibility allows users to tailor the analysis of events to specific requirements.
In this section, we will present the components of our multi-step scheme and options available to the user, discussing both the novel and traditional approaches implemented in \pkg{dupin}.

\subsection{Detection Scheme}\label{sec:scheme}
\pkg{dupin}'s detection scheme is based on CPD~\cite{aminikhanghahi.cook2017,truong.etal2018}.
CPD seeks to assign a set of points, $K$, where a signal, $S : \{s_0, s_1, \ldots, s_N\}$, undergoes a change.
We denote a single point in a signal (e.g.\ $s_i$) as a frame.
We define two operators - indexing and slicing - that act on the signal.
The indexing operator returns a single value such that $S[i] = s_i$.
The slicing operator produces a sub-signal using a semi-closed interval $S[i,j): \{s_i, s_{i+1}, \ldots, s_{j-1}\}$.
Using the slicing operation, we can see that the set of points $K$ encompass $|K| + 1$ sub-signals between $[k_{i}, k_{i+1})$, where $k_0 = 0$ and $k_{|K|}$ is the last frame in the signal.
The first and last change points, $k_0$ and $k_{|K|}$ are trivial, so the number of change points is often written as $|K| - 2$ instead of  $|K|$.
We adopt this convention from this point on in the paper.

\begin{figure*}[htpb]
    \centering
    \includegraphics[width=\linewidth]{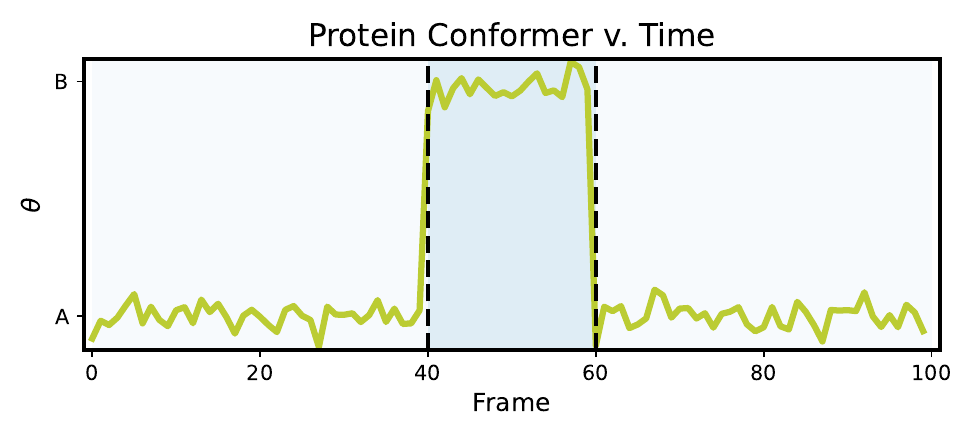}
    \caption{%
        An example detection of a protein transitioning between conformers A and B.
        At frame 40, the protein goes from conformer A to B and back to A at frame 60.
        The background colors and dashed black line indicate these two change points, and the corresponding three sub-signals are $S[0,40), S[40, 60), S[60, 100)$.
        $\theta$ represents a fictitious order parameter along which a structural change happens.
    }%
    \label{fig:subsignals}
\end{figure*}

As an example of CPD, imagine a 100-frame trajectory of a protein with two conformers, A and B, and a signal $S$ comprising data representing the conformation of the protein in each frame.
If the protein changes conformation at frames 40 and 60, then the signal $S$ can be sliced into three sub-signals (Figure~\ref{fig:subsignals}): $S[0,40), S[40,60),$ and $S[60,100)$.
In this case, there are two change points $K = \{40, 60\}$.
This is because the random fluctuations (the fluctuations of a (metastable) equilibrium system around the mean) should be ignored.

We employ a class of CPD algorithms that use the paradigm of optimization and loss/cost functions to select change points from $S$~\cite{truong.etal2018}.
A loss or cost function penalizes some notion of error, which is then optimized to minimize the \textit{cost}.
We find a set of change points $K$ of size $n$, such that choice of each single change point $k \in K$ minimizes the cost function $C$,

\begin{equation}\label{eq:cpd-opt}
    K = \argmin_k \sum_{i=1}^{n+1}{C(S[k_{i-1}, k_i))},
\end{equation}

\noindent where $S[k_{i-1}, k_i)$ is the partition of the signal $S$ between potential change points $k_{i-1}$ and $k_i$.
Notice that the number of change points expected, $n$, in the signal must be known in advance.
A more detailed exploration of this issue is provided in the subsequent discussion.
Further elaboration of this class of CPD algorithms is given in~\suppref{sec:cpd}.

\subsubsection{Overview of Method}
\begin{figure*}[htpb]%
    \centering
    \includegraphics[width=\linewidth]{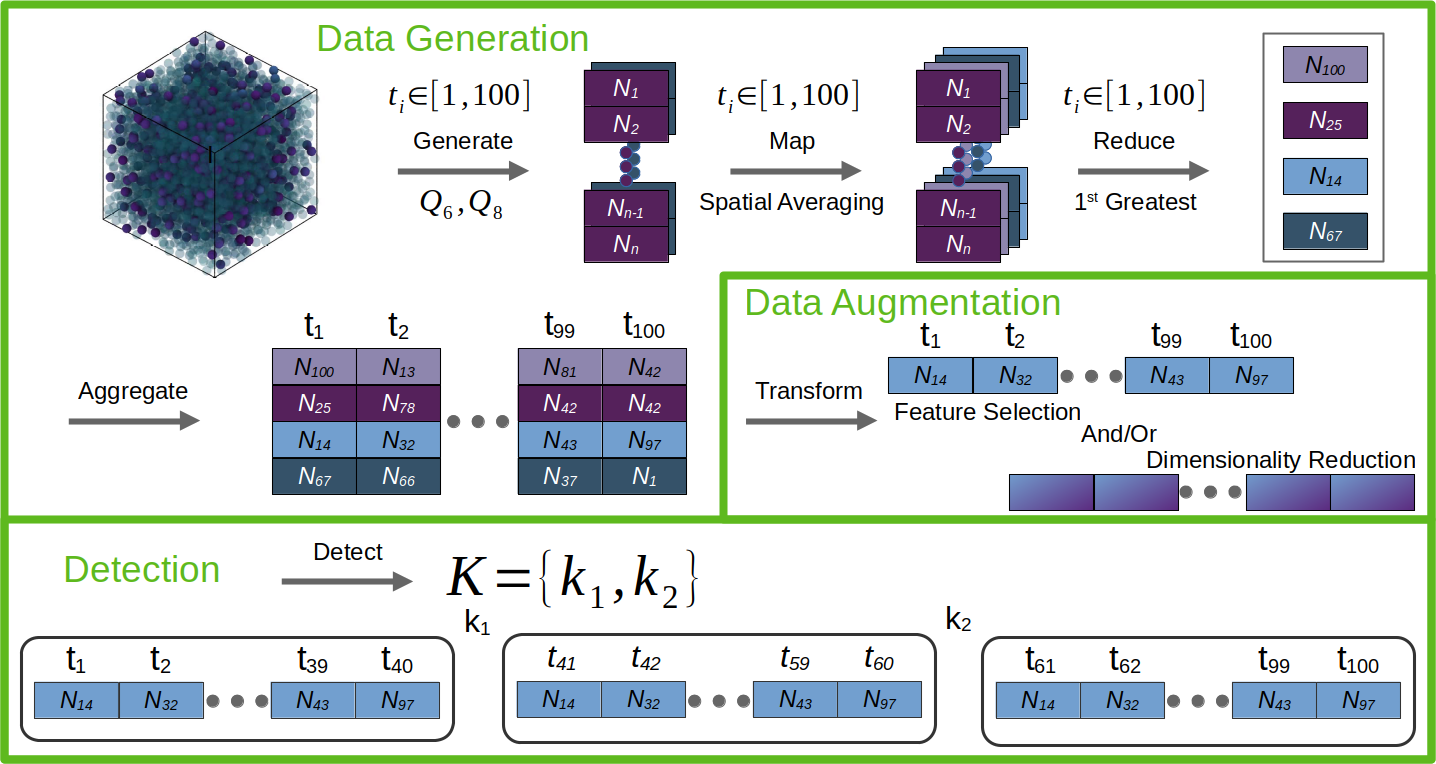}
    \caption{%
    An illustration of the typical pipeline for event detection in a particle trajectory.
    We use the same numbers as the protein conformer example (change points are 40 and 60 and the total frames are 100).
    Arrows indicate steps and small colored rectangles represent individual per-particle feature values generated from the trajectory.
    Green boxes separate steps into three general stages: data collection, data augmentation, and detection.
    Data is \textit{generated} from the frame and \textit{mapped} to a new distribution and to itself (note the replication of the original features after mapping).
    The distributions are then \textit{reduced} into one scalar feature each.
    For the \textit{generate}, \textit{map} and \textit{reduce} steps underneath the arrows, we provide potential functions/applications of the step.
    The features are then \textit{aggregated} across frames, and \textit{transformed} with either feature selection or dimensionality reduction.
    Finally, the change points as well as the number of change points are \textit{detected}.
    The \textit{transform} step is not required, and the \textit{aggregate} step could immediately precede the final step, \textit{detect}.
    The figure assumes per-particle features only.
    For global features, the \textit{generate} step immediately precedes the \textit{aggregate} step --- they could also optionally be \textit{mapped} first.
    }%
    \label{fig:flow-chart}
\end{figure*}

To develop a generic protocol for event detection in molecular trajectories, we must answer three fundamental questions (i) how to generate a signal from a trajectory, (ii) what cost function(s) best partition the trajectory into sub-signals and (iii) how to determine the correct number of change points in a signal to ensure detection of all events.
To address these problems, \pkg{dupin} uses a novel approach in which different steps are grouped into stages (see Figure~\ref{fig:flow-chart}).
We group the steps into three separate stages: data collection, data augmentation, and detection.
The data collection stage includes the \textit{generate}, \textit{map}, \textit{reduce} and \textit{aggregate} steps.
The data augmentation stage includes the \textit{transform} step.
The detection stage includes the \textit{detect} step.
The \textit{generate, aggregate} and \textit{detect} steps within the data generation and detection stages are always required.
The \textit{reduce} steps can be required or optional depending on the data generated while the \textit{map} and \textit{transform} steps are always optional.
The schematic of a typical pipeline for event detection in \pkg{dupin} is presented in Figure~\ref{fig:flow-chart}.

\subsubsection{Data Collection}%
\label{ssec:scheme-collection}
We begin by constructing feature vectors for every frame up to a total number of frames $N_\mathrm{frames}$.
A feature vector is a set of $n$ features used to describe a data point in an $n$-dimensional space.
Order parameters (MSM, local density, etc.) are examples of features.
For our case, we refer to features that describe aspects of a given point/time in a trajectory.
We combine the feature vectors into a signal $S$, which is a matrix of size $N_\mathrm{frames} \times N_\mathrm{features}$.
Each matrix element $s_{ij}$ in $S$ contains the value of one of the features in one of the frames of the trajectory.
We assume that any change in a molecular system can be adequately described by a such a signal.
Changes in the system are thus indicated by changes in the feature vectors over time/frames.

The \textit{generate} step requires us to choose the class of features that will be computed for every frame.
Once the feature data is generated, non-scalar or vector features (e.g.\ per-particle quantities such as Steinhardt order parameters~\cite{steinhardt.etal1983}) must first be reduced.
Reduction takes a vector feature and converts it through a variety of reducers (functions) to a finite number of scalar features representative of the distribution.
Scalar features are never reduced (they are already scalar).
The \textit{reduce} step cannot be used on global properties (scalar features) such as system potential energy, but is required for per-particle or high dimensional properties (vector features) like the per-particle potential energy.
Examples of scalar features resulting from a \textit{reduce} step include the maximum value, minimum value, mean, mode, median, range, $n$-th greatest, $n$-th least, etc.
\pkg{dupin} offers several novel reducers that are not commonly used in literature, such as the $n$-th greatest and least reducers.
Such reducers aren't useful in settings the descriptors are usually used, such as description of global structure, but are the appropriate choice for detecting transitions in trajectories.
This is because initial transition events occur at tails (minima or maxima) of the descriptor distributions.

As an example, consider the utility of the $n$-th greatest and the $n$-th least reducers in the context of a study of crystallization.
The process of homogeneous nucleation and growth of a crystal from a metastable fluid phase starts when local fluctuations cause a group of neighboring particles to form an ordered cluster.
If this cluster reaches or exceeds a critical size, it begins to grow, leading to the production of the crystal phase~\cite{karthika.etal2016}.
This process is known as classical nucleation theory.
Ideally, CPD would have access to information on the initial fluctuations conspiring to produce the cluster as well as the cluster's subsequent growth.
In the following analysis, we assume (i) we have already constructed or defined features that adequately distinguish the solid and fluid phases, and (ii) the initial fluctuation is small compared to the system size.
To detect the nucleation event, we must use reducers closer to the extrema such as the 1st and 10th greatest or least values (e.g.\ for local densities).
Selecting near the extrema is necessary as most particles are fluid when a nucleus forms and selecting reductions that average over the distribution or select near the mean will not register a nucleation event.
On the other hand, as the nucleus grows to 10 then 20 then 50 then 100 particles, more particles assume the approximate feature vector of the crystal.
Thus, the 100th greatest or least reducers would capture the growth of a cluster up to a size of 100 particles.

While never required, vector features can be mapped to other distributions before aggregating.
An example of a useful mapping is spatial averaging of a feature over its neighbors as is sometimes done with Steinhardt order parameters~\cite{lechner.dellago2008}.

After a set of scalar features describing a single frame of the trajectory is generated, the process repeats across all trajectory frames and the results are \textit{aggregated }into a single multi-dimensional, time-dependent signal.
This signal can be sent directly to the \textit{detect} step or to the \textit{transform} step.
The \textit{transformation} step can involve any combination of three tasks: feature selection, dimensionality reduction and signal filtering.
Next, we describe each of these data augmentation tasks as they pertain to \pkg{dupin}.

\subsubsection{Data Augmentation}%
\label{ssec:scheme-augmentation}
After features have been reduced, the number of features may easily be in the hundreds.
In particle trajectories, each feature incurs noise due to thermal fluctuations --- the same fluctuations involved in, e.g., nucleation and growth events in crystallization.
As a result, this thermal noise increases the chance of spurious or undetected events for two related reasons.
First, one or more features may fluctuate enough to be mistaken as an event.
If we assume a baseline chance for such a fluctuation to be 1\%, then a signal of 200 features has an 86.6\% chance of recording such a spurious event.
Second, given $n$ features, if the event appears in only one feature, then as $n \to \infty$ the relative reduction in cost to fitting to that one feature decreases drastically.
Assuming comparable noise in each feature, then for $n$ features each feature contributes $1 / n$ to the cost.
For 200 features, the reduction in cost for fitting to a single feature's change point is at most 0.5\% percent of the original cost.

To prevent the noise in high feature dimensions from leading to poor event detection, data augmentation through feature selection or dimensionality reduction can improve performance.
The amount of improvement depends on the dimension of the feature set, noise level, and other characteristics of the signal.
Furthermore, data augmentation drastically improves computational performance of the event detection scheme without diminishing detection performance.
The cost of the detection algorithm used in this paper is linear in the dimensionality of the feature set; that is, doubling the number of features reduces the performance by half.

\pkg{dupin} currently implements two different approaches to feature selection: (a) ``mean-shift'' filtering and (b) feature correlation.
The main idea of mean-shift filtering is to compare two distributions and determine if they have a significantly different means from each other.
For our purposes the distributions considered are the distributions of features in the system at the beginning and the end of the signal.
A significant change in the mean of these features suggests that the feature effectively captures an event within the system.
To perform this comparison, we analyze the means and standard deviations of both distributions to assess whether the means are equivalent.
Based on this comparison and a predefined sensitivity parameter, we then make a \textit{yes/no} decision about whether to include the descriptor for event detection.
For more information see \suppref{sec:feature-selection}.
Note that the mean-shift filtering is different from a mean-shift cost function detection.

Feature selection via feature correlation is done through spectral clustering~\cite{andrewy.ng.etal2002}, where the similarity matrix is computed based on the correlation matrix of the signal.
A pre-determined number of features from each cluster is taken based on a provided feature importance score or is randomly selected from a cluster.
More information on these methods for feature selection can be found in~\suppref{sec:feature-selection}.
Other feature selection methods such as forward selection or backward selection seamlessly interoperate with \pkg{dupin} and can be used as well.
Both methods can be found in scikit-learn~\cite{pedregosa.etal2011}.

\begin{figure*}%
    \centering
    \includegraphics[width=\textwidth]{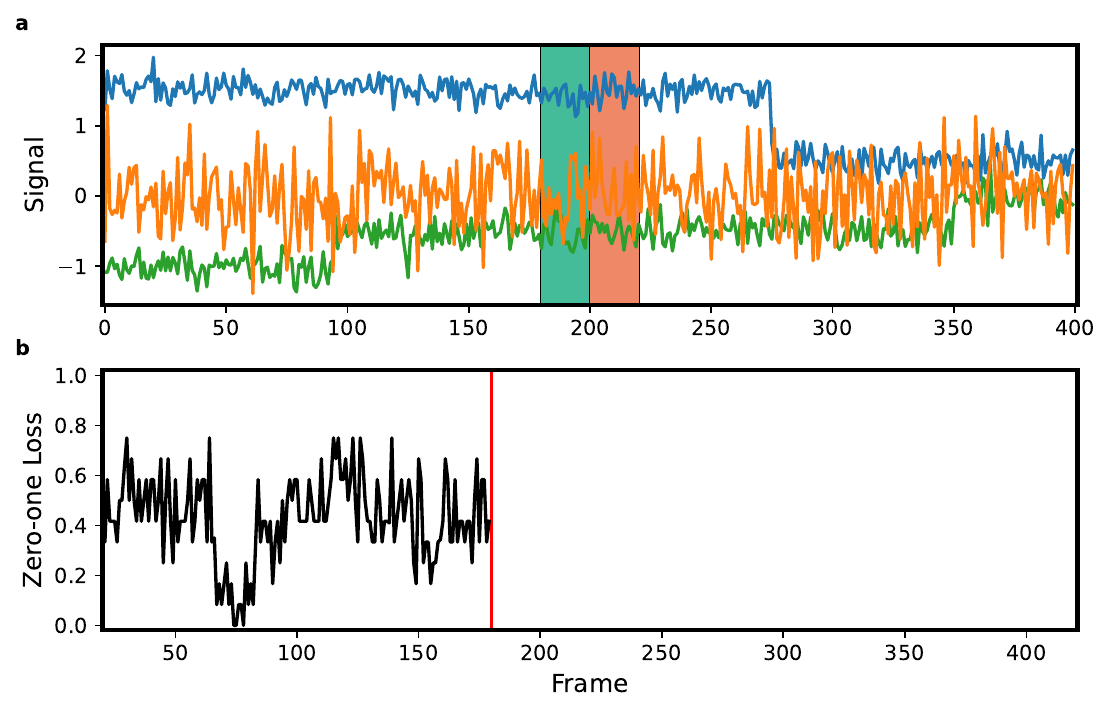}
    \caption{%
        An example highlighting the use of classifiers to reduce a three-dimensional signal to one dimension.
        The process was stopped with a window center frame of 200 for instructive purposes.
        (a) A plot of the three-dimensional signal.
        Change points are located at 75, 275 and 350.
        The box centered at frame 200 represents the two window halves: $[190, 200)$ and $[200,210)$.
        (b) The zero-one loss from the start of the signal to the current window.
        Notice the drop towards zero loss near frame 75, which corresponds to a mean-shift in (a) (blue line).
        }%
    \label{fig:ml-window}
\end{figure*}

For dimensionality reduction, \pkg{dupin} implements a novel machine learning (ML) classifier to reduce the signal to a single dimension based on local signal similarity; this approach is a  variation of the approach found in Reference~\cite{hido.etal2008}.
Signal reduction is accomplished by taking a sliding window across the original signal and using an ensemble (collection) of weak learners to determine the similarity of the window's left and right halves.
A weak learner is one with limited ability to discriminate between classes (e.g.\ a decision stump~\cite{sleeman.edwards1992} that classifies based on a single \textit{yes/no} condition).
Thus, weak learners cannot distinguish window halves based solely on noise due to their limited discrimination ability.
This limitation is a desirable property as we only want the classifier to discriminate on \textit{significant} differences between window halves.
Each learner in the ensemble is trained and tested on different data within the window via a stratified shuffle split (from scikit-learn), which reduces the chance that a bad data partition will decrease the test loss.
To determine the local signal similarity, we label each frame in the left window half with the class label zero and each frame in the right window half with class label one.
The classifiers within the ensemble are then trained on a subset of the data and tested on the remaining data.
The testing loss (e.g.\ the zero-one loss) is used as a metric of dissimilarity.
When there is no event, the classifiers should have nothing but random noise to train on, resulting in the classifier being wrong on test data $\sim$50\% of the time.
However, when an event occurs within the window, a classifier can train on differences between the window halves and accuracy tends towards 100\% (i.e.\ 0.0 loss).
For each window position, we take the average loss as our one-dimensional signal (here, the zero-one loss).
Figure~\ref{fig:ml-window} shows a graphical depiction of the dimensionality reduction implemented in \pkg{dupin}.

The ML classifier has several hyperparameters, including the window size, number of classifiers (decision stumps) and the  percentage of the data used for training.
The window size must be chosen based on the number of samples (trajectory size).
In general, bigger is better unless it leads to only a few windows or if it's much larger relative to the length of the transition.
The choice of dumping frequency can also have important implications for the window size.
If the dumping frequency of the feature is smaller compared to its relaxation time, the window size should be large enough that it contains several uncorrelated samples.
This way we ensure that training is done on at least partially uncorrelated samples (random fluctuations).
The number of classifiers determines smoothness of the error.
Higher numbers naturally smooth the error across a trajectory and are generally preferable.
The highest possible number of classifiers should be half of the non-repeating combinations of training and test set.
This limit ensures that for small window sizes enough classifiers are given while being relatively confident that most window data splits will be unique.
The test size is the percentage of the data used for testing.
Larger test sizes provide better representation of the true representation.
A consequence of this is that larger test sizes result in better error estimates and increased smoothness of the error.
In general, test size should be smaller than half but not too small, to ensure comparison to correlated samples by chance is avoided.

Other schemes and algorithms for dimensionality reduction such as principal component analysis (PCA), uniform manifold approximation and projection (UMAP)~\cite{mcinnes.etal2020}, etc.\ can be used to reduce the number of features into a few information-dense dimensions as well.

Signal filtering is commonly used for smoothing a signal and is similar to the \textit{mapping} step in the data generation stage.
Here, the completed signal is transformed along the temporal dimension into a new signal.
Signal filtering is commonly used for smoothing a signal.
\pkg{dupin} has a rolling mean signal filter, which smooths noise by averaging the signal across neighboring frames.
Other signal filters are available in packages like SciPy and can be used easily with \pkg{dupin}.

\subsubsection{Detection}%
\label{ssec:scheme-detection}
The final stage in \pkg{dupin}'s pipeline is event \textit{detection}.
To detect local events, two cost functions are available for use in \pkg{dupin}.
Both are based on piecewise linear fits of time versus signal features, and both have increased cost when this fit has a higher summed $p$-norm error $|y - f(x)|_p$.
In the examples in Section~\ref{sec:systems}, we use the summed and square rooted 2--norm.

The first cost function, $C_1$, in \pkg{dupin} computes the $p$-norm loss of the piecewise, least-squared linear fit of each feature as a function of time.
We refer to this approach as linear cost function approach.
This procedure minimizes the $p$-norm via a least-squared fit on the slope $m$ and intercept $b$ for the given sub-signal $S[i,j)$.
To prevent units or feature magnitudes from contributing to faulty detection, we map all features to the unit square independently so that the range for each feature is $[0,1]$.
After mapping to the unit square, $C_1$ is given by:

\begin{equation}\label{eq:cost-linear}
    C_1(S[i,j)) = \min\limits_{m, b}{\sum_{x=i}^{j}{|S[x] - (m x + b)|_{p}}},
\end{equation}

\noindent where $||_{p}$ is the $p$-norm, $S[x]$ is the signal value at frame $x$, and $m x + b$ is the linear fit of the signal where $m$ is the slope of the linear fit and $b$ is the y-intercept.

The second cost function, $C_2$, is similar to the first, but the linear fit is determined simply by drawing a line from the beginning point to the end point of the sub-signal, without any fitting.
We refer to this approach as biased cost function approach.
As a result, this cost function is more sensitive to sudden shifts and changes in slope in the signal compared to $C_1$.
The $C_2$ cost function is given by:

\begin{align}\label{eq:cost-biased}%
    C_2(S[i,j)) &= \sum_{x=i}^{j}{|S[x] - (m x + b)|_{p}} \\
    m &= \frac{S[j] - S[i]}{j - i} \nonumber \\
    b &= S[i] - m i \nonumber,
\end{align}

The only optimizable parameters in Equation~\ref{eq:cost-biased} are the locations of the change points.
In cases where greater sensitivity is desired, $C_2$ is a viable alternative.

\pkg{dupin} can also detect events by using a simple mean-shift cost function detection approach.
In this approach, a mean-shift cost function is computed instead of fitting linear functions to the signal.
Such approach is especially effective when the ML classifier is used for the data augmentation step.
Note that the mean-shift cost function detection is different from mean-shift filtering.
For more details on this approach including the cost function definition see \suppref{sec:mscf}.

Any detection algorithm can be used in \pkg{dupin}.
For this work we use the Python package \pkg{ruptures}~\cite{truong.etal2018}, which implements various algorithms for solving the optimization problem already posed (see~\suppref{sec:cpd} for more information).
\pkg{dupin} provides the implemented cost functions above to the detection class from \pkg{ruptures} (we use the dynamic programming solver in this work).
Using \pkg{ruptures}, we find the optimal change points for a given number of change points.

After detection, we still have one more problem to solve: finding the optimum number of change points.
To do this, we find an elbow in the total cost function as a function of the number of change points $n$.
The elbow is defined as the point of maximum curvature, but can be expanded to discrete points.
\pkg{dupin} can use any elbow detection method that works with discrete points.
We choose for this work the kneedle algorithm~\cite{satopaa.etal2011} found in the \pkg{kneed} Python package; kneedle behaves well with the test systems studied and provides a hyperparameter allowing for sensitivity control in elbow detection.
The sensitivity parameter also means the algorithm can return no elbow, allowing \pkg{dupin} to select zero change points as the correct CPD.
More details on the kneedle algorithm and its usage in \pkg{dupin} can be found in~\suppref{sec:kneedle}.

\subsection{Online Detection}%
\label{ssec:online}

Another novelty enabled by \pkg{dupin} is the ability to perform online detection of events with approaches outlined above.
Online detection is the process of detecting events while the particle trajectories are still being generated.
This approach is useful in cases where responses to events are desired or when the generated trajectory is too large to be stored in memory.
In online detection, the \textit{detection} step is applied on-the-fly to the set of frames generated up until the moment of detection.
This set of frames can extend to the beginning of the trajectory or be a sliding window of the last $N$ generated frames.
To use \pkg{dupin} online, a predefined set of features is calculated on-the-fly at the desired frequency leading to a continuous application of \textit{generate} to  \textit{aggregate} steps.
If using a sliding window rather than the whole trajectory, the order parameters are placed into a ``first-in first-out'' (FIFO) queue, so that the sliding window moves with the time evolution of the data.
Sliding window is the preferred approach for online detection, due to improved performance and easier CPD setup for detection.
Using the queue approach allows CPD to be run on only part of the trajectory, which dramatically speeds up detection, making it more viable for online use.
Each time a new frame is added to the signal, CPD can be run on the current data in the queue.
The algorithm works best when the window size is commensurate with the size of a single event.
Either way, \pkg{dupin} should be run with the assumption that we are expecting to detect up to one event.
However, in practice we run CPD up to a change point set size of approximately four as any elbow detection scheme will need some points beyond the elbow to detect it.
To detect multiple events, the queue should be cleared after detecting an event.
If this is not done that event will continue to be detected in consecutive runs, leading to undesirable behavior.

\section{Example Applications}%
\label{sec:systems}
We now demonstrate \pkg{dupin}'s event detection scheme step-by-step using \pkg{dupin} for two example systems to highlight its usefulness and versatility.
All trajectory data was produced by molecular dynamics (MD) simulations run using HOOMD-blue~\cite{anderson.etal2020,anderson.etal2016,butler.etal2020}.
Feature vector construction (\textit{generation}) was carried out using \pkg{freud}~\cite{ramasubramani.etal2020,dice.etal2019} in conjunction with \pkg{dupin}.
The data was organized and managed using the \pkg{signac} framework~\cite{adorf.etal2018b,ramasubramani.etal2018,dice.etal2021}.
System visualization was performed using OVITO~\cite{stukowski2009}.

\subsection{Binary Lennard-Jones system}
\label{ssec:binary-lj}

\begin{figure*}%
    \centering
    \includegraphics[width=\linewidth]{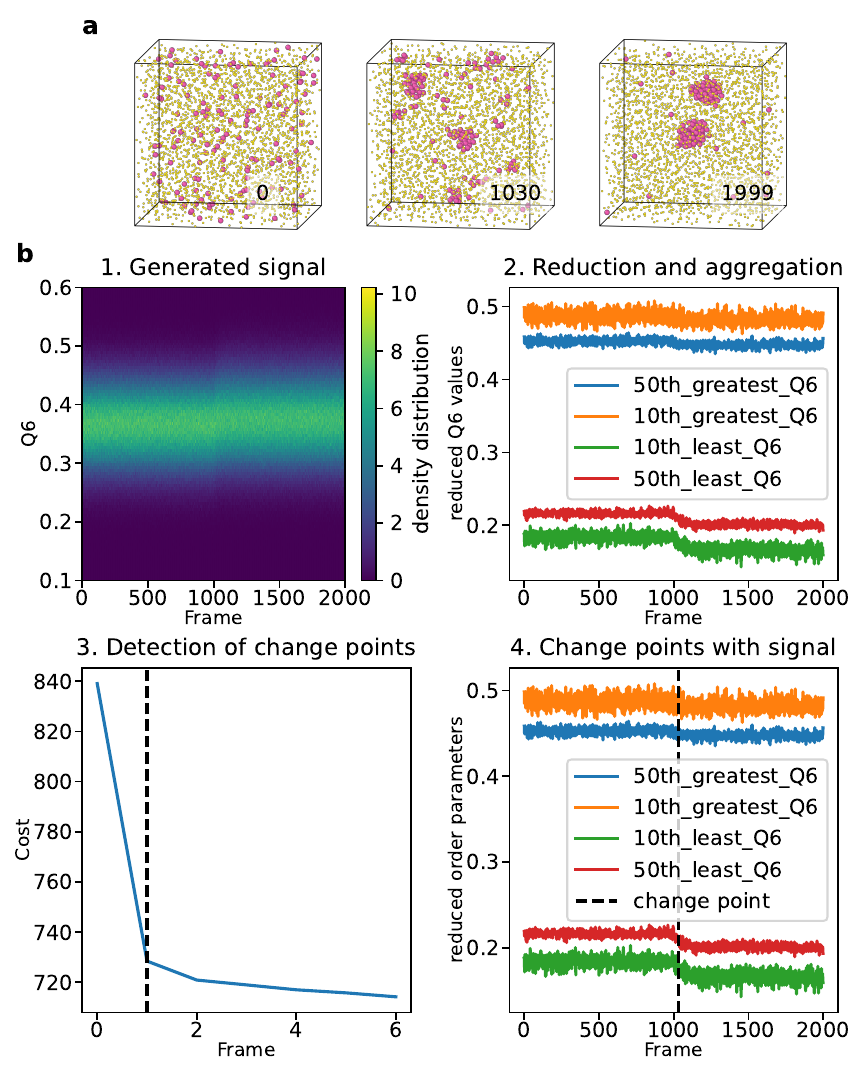}
    \caption{%
        Change point detection applied to system of binary LJ particles.
        (a) Simulation snapshots with associated frame number (lower right).
        The middle image shows the system at the detected change point.
        Particles are colored according to their type (A - yellow; B - pink).
        (b) 1. The result of the signal generation step is represented as a density distribution of $Q_6$ for each frame of the trajectory.
        2. After applying reduction and aggregation we are left with final signal on which the detection will be performed.
        3. The cost plot used to compute the optimal number of change points using elbow detection (black dashed line) for linear change point detection.
        4. Resulting change point (black dashed line) plotted on the signal.
    }
    \label{fig:LJ}
\end{figure*}

Our first example is a binary Lennard-Jones (LJ) system containing particles of type A and type B.
This example demonstrates that \pkg{dupin} can be used to detect change points even if only a small portion of the system undergoes a transition.
The MD simulations were performed in canonical (NVT) ensemble at temperature $T=1.5$, density $\rho=0.8$.
System contained a total of $N_p=2916$ particles of which 145 $(5\%$ of the system) were particles of type B, and the rest were particles of type A.
The simulation starts off with all particles interacting with the same LJ potential ($\epsilon=1$, $\sigma=1$ with cutoff $r_{\mathrm{cutoff}}=2.5$).
To prepare the system for production run we first place the particles on the simple cubic lattice, melt it, and equilibrate the system.
Figure \ref{fig:LJ} (a) left shows the randomized and equilibrated system at the beginning of the production run.
To generate the trajectory of point cloud data for detection we saved the system snapshot every 1000 time steps.
The production part was run in two phases.
In the first phase we ran for 1 million time steps.
Next, the interaction between particles of type B is changed in such a way that LJ interaction potential parameters are changed to $\epsilon=2.5$, $\sigma=0.5$ without changing the cutoff.
In second phase we ran for another million time steps.
The production run generated $N_f=2000$ frames in total.

We set up \pkg{dupin}'s detection pipeline to compute the Steinhardt order parameter for $l=6$ ($Q_6$) with 12 nearest neighbors for each particle in the system as our \textit{generate} step.
This step results in $N_p \times N_f$ values of $Q_6$ in total.
We represent this step as a distribution of $Q_6$ as a function of time in Figure \ref{fig:LJ} (b) 1.
The \textit{map} step is skipped and the \textit{reduce} step is applied to find the 10th, and 50th greatest and least values of the $Q_6$ distribution.
The features produced by the reduce step are kept and aggregated into the final signal which contains $N_f \times 4$ values in total (Figure \ref{fig:LJ} (b) 2.).
The \textit{transform} step was skipped because the dimensionality of the signal is already low.
To \textit{detect} change points in the trajectory we used the $C_1$ linear cost function.
Elbow detection was used to compute the optimal number of change points (Figure \ref{fig:LJ} (b) 3.).

The detection scheme employed resulted in a single change point detected at frame 1030 (Figure \ref{fig:LJ} (a) middle).
The detected change point corresponds to a frame occurring 30000 time steps after the interaction potential between particles of type B was changed.
The employed change in the potential between particles of type B caused particles of type B to aggregate into clusters due to much stronger interaction.
At frame 1030 (Figure \ref{fig:LJ} (a) middle) we can clearly see that several smaller clusters of particles of type B have already formed.
We can roughly estimate from Figure \ref{fig:LJ} (b) 2. that the transition started at frame 1000, when the potential was changed and ended at around frame 1100.
This places the detected change point squarely in the middle of the transition.
The snapshot shown in Figure \ref{fig:LJ} (a) right, corresponds to the simulated system at the end of the simulation.
We can see that that most of particles of type B have aggregated into two larger clusters by the end of the transition.
This example showcases how \pkg{dupin}'s change point detection capabilities even if events are undergone by a small part of the system.

\subsection{Nucleation and growth of a binary crystal of particles}
\label{ssec:binary}
Our second example is a binary system of point particles interacting via the Mie potential~\cite{mie1903} ($n=50$, $m=25$) with a size ratio of 0.55.
The simulation was run using MD for 36.8 million time steps in the isothermal isobaric ensemble (temperature $T=0.35$, pressure $P=0.052$, number of particles $N_p=27,000$, in reduced units) to simulate the solidification of a liquid into a crystal by homogeneous nucleation and growth.
The simulation was initialized in a NaCl lattice with random placement of the two species, which quickly dissolves upon thermalizing at slightly higher temperature prior to a quench to the target $T$.

\begin{figure*}%
    \centering
    \includegraphics[width=\linewidth]{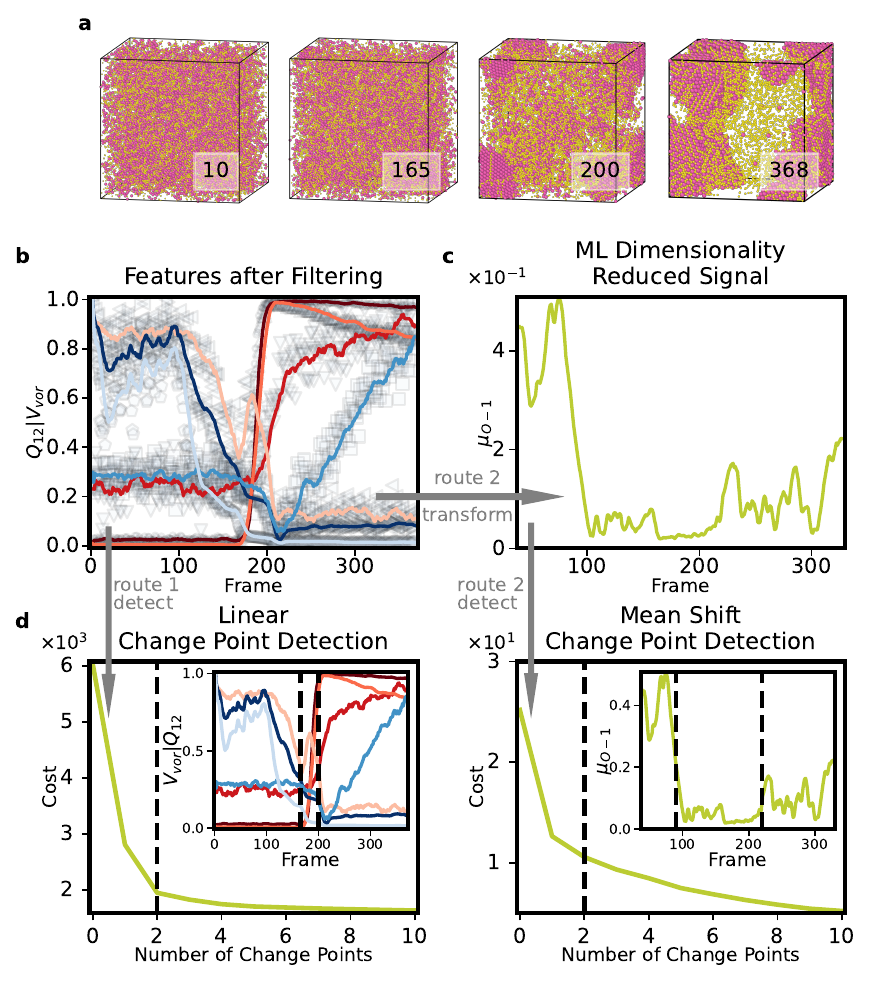}
    \caption{%
        Change point detection applied to system of binary Mie particles.
        (a) Simulation snapshots with associated frame number (lower right).
        The two middle images show the system at the two change points detected using linear change point detection.
        Particles are colored according to their type.
        (b) Plot of some features obtained after the transform step (100th greatest and least MSM $l=12$ (left) and Voronoi polyhedra volumes (right)).
        Solid lines are the (rolling mean) smoothed features and the translucent points are the actual data points used for calculations.
        Gray arrows and text indicate two different change point detection routes applied.
        (c) The zero-one loss function computed using the ML classifier for dimensionality reduction applied to features from transform step (b).
        (d) The two main figures show the cost plots used to compute the optimal number of change points with elbow detection (black dashed line) for linear change point detection (left) and mean-shift change point detection (right).
        The insets show the change point frames (black dashed line) on the signals used to compute the change points.
    }
    \label{fig:binary}
\end{figure*}

To \textit{generate} the signal we compute the Voronoi polyhedron volume~\cite{voronoi1908,voronoi1908a} and MSM~\cite{mickel.etal2013} for spherical harmonics $l = 2, 4, 6, 8, 10, 12$ for each particle.
For each feature, we \textit{map} to two distributions --- itself (no transformation) and the Voronoi tessellation neighbor average~\cite{lechner.dellago2008}.
We then \textit{reduce} each distribution (raw and averaged) to six features: the 1st, 10th, and 100th greatest and least values.
After \textit{reducing}, the MSM for each $l$ produces twelve features, six from the raw distribution and six from spatial averaging.
Following \textit{aggregation}, we \textit{transform} the signal via feature selection through a mean-shift filter with sensitivity of $10e{-4}$.
After the mean-shift filter is applied we are left with the signal containing several features which change significantly during the length of the trajectory.
Some of the features that pass the mean-shift filter are shown in Figure \ref{fig:binary} (b).
Next, we showcase two different detection approaches.
In the first detection route, we take the filtered signal and \textit{detect} the change points using $C_1$ (linear cost function) with \pkg{rupture}'s dynamic programming algorithm, using \pkg{kneed} for elbow detection with a sensitivity of 1, for $|K| \in [1, 10]$.
For the second detection route, we take the filtered signal and we first \textit{transform} the signal again by applying the ML classifier dimensionality reduction (window size 80) introduced in Subsection~\ref{ssec:scheme-augmentation}.
We used 200 decision stumps (decision trees of depth one) on features selected by the mean-shift filter to classify windows halves.
The test set size used was 0.4, which means that 40\% of the data in the window was not used for fitting.
The zero-one loss is then smoothed over the neighboring three errors on each side with the mean signal filter.
The new signal obtained from the ML classifier dimensionality reduction route was then used to \textit{detect} events using a $L_1$ mean-shift cost function~\cite{jandhyala.etal2013}.

The first detection route (Figure \ref{fig:binary} (d) left) which \textit{detects} using $C_1$ linear cost function detection results in two change points.
The linear cost function detection was dominated by a region of sharp change in several properties $\approx$160--200 during crystallization.
We note that the continuing shift of some Voronoi polyhedra volumes or MSMs at the end of the simulation is roughly linear, meaning $C_1$ does not penalize grouping them into a single sub-signal.

On the other hand, the second detection route (Figure \ref{fig:binary} (c) and (d) right) in which we first add another \textit{transform} step using the ML classifier for dimensionality reduction (Figure \ref{fig:binary} (c)) and then \textit{detect} using mean-shift cost function (Figure \ref{fig:binary} (d) right) results in a much larger transition event window (i.e.\ the change points are farther apart).
Such a partitioning is expected as the dimensionality reduction scheme picks up on \textit{any} deviation across the window.
As a consequence, mean-shift cost function route detected change points at frames 90 and 210, in contrast to the linear cost function route.
We note that the reason for the change point locations can be seen in Figure~\ref{fig:binary} (c) where the smoothed average zero-one loss is plotted.
The system is still undergoing structural changes at frame 210, indicated by the value of average loss which while higher than 0.0 is lower than the expected value outside a transition of 0.5.
This behavior is explained by the observed trends in some Voronoi polyhedra volumes and MSMs, which are still not in equilibrium (flat) after frame 210.
The final slice of the trajectory (frame 210 till the end) is thus associated with a new phase of the ongoing transition, which is not finished by the end of the trajectory as indicated by the average loss value.

%\textcolor{red}{
%In the nucleation event studied in this example, only part of the system undergoes the nucleation event.
%This showcases that \pkg{dupin} is capable of detecting events in a system where only a fraction of the particles are involved in the event.
%}
These same schemes can be followed to apply \pkg{dupin} to simulations of molecules, nanoparticles, colloids, polymers, or other "particle"-based systems that generate particle positions (and possibly orientations) as a function of time, making \pkg{dupin} highly extensible and generalizable.

\subsection{Collapse of polymer in poor solvent}
Our third example involves detecting the collapse of an isolated polymer chain in an implicit solvent following a change from good to poor solvent using \pkg{dupin}.
We simulated a polymer comprised of $5000$ connected polymer beads interacting via a Lennard-Jones pairwise interaction, with bonds between beads modeled by a simple harmonic potential $U = - 0.5k(r - r_0)$.
The MD simulation was performed in the canonical ensemble (number of particles $N=5,000$, system volume $V=10^6$, and temperature $T=1$).
Following equilibration (1.2 million steps) in good solvent, the simulation was run for 2.5 million steps, where the $\varepsilon$ of the Lennard-Jones potential was increased to mimic a change in the Flory $\chi$ parameter to a poor solvent.
Over the 2.5 million steps, $\varepsilon$ was increased from 0.1 to 1.25 over 50 intervals of 30,000 steps, running at the final $\varepsilon$ for 1 million steps.
Increasing the $\varepsilon$ parameter causes the polymer beads to start aggregating.
Several stages of aggregation are observed in which the number of lobes decreases in stages until final metastable configuration of two lobes is obtained.

We \textit{generate} the signal using the polymer end-to-end distance, the local density defined as Voronoi polytope volume, and the number of clusters of neighboring beads using \pkg{freud}'s clustering algorithm ($r=1.2$).
Because Voronoi polytope volume is the only non-scalar feature, we \textit{reduce} only it.
We \textit{reduce} the volumes to six features: the 10st, 100th, and 1,000th greatest and least values.
We do not \textit{transform} the data due to low dimensionality.
Finally, for offline detection, we \textit{detect} the change points using $C_1$ with \pkg{rupture}'s dynamic programming algorithm, using \pkg{kneed} for elbow detection with a sensitivity of 1, for $|K| \in [1, 10]$.
We also perform online \textit{detection} using the sliding window approach with a window size of 50.
All steps up to \textit{detection} are identical though the signal is fed to the detector as a stream rather than simultaneously.
The detector used for the online \textit{detection} uses the same cost function, algorithm, and elbow detection, but only goes to $|K| = 6$.
The cost function gets computed on-the-fly for each frame in the window, taking into account previous frames in the window.
We clear the window whenever an event is detected.

\begin{figure*}[htpb]
    \centering
    \includegraphics[width=\textwidth]{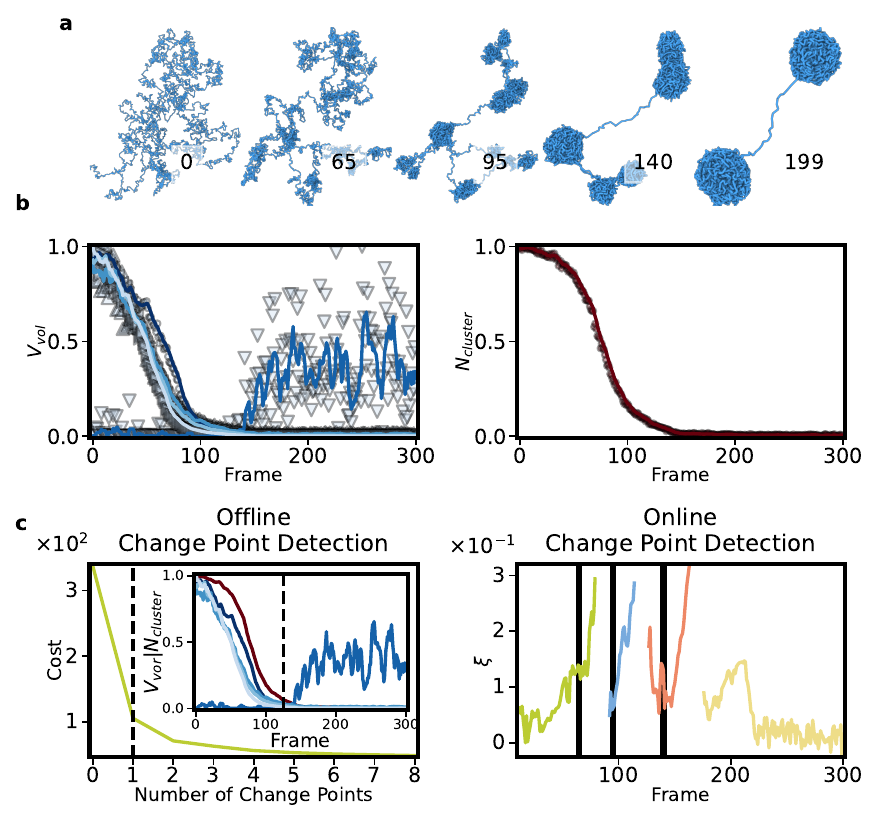}
    \caption{%
        Change point detection applied to polymer trajectory.
        (a) Images corresponding to the beginning of the simulation, online change points, and end of the simulation.
        (b) Plots of features for detection: Voronoi polytope volumes (left) and the number of clusters (right).
        Solid lines are the (rolling mean) smoothed features and the translucent points are the data points on which all calculations were done.
        (c) Detection of change points using the offline (left) and online approach (right).
        Offline plot (left) of the cost associated with the optimum $n$ change points computed on all features.
        The inset for the offline plot (left) shows the features from (b) with the computed change points (black dashed line).
        Online detection (right): Window cost proxy function (see main text) for online detection and associated change points (black lines).
    }
    \label{fig:poly}
\end{figure*}

Figure~\ref{fig:poly} shows the CPD analysis for this system.
The collapse of the polymer due to poor solvent can be seen in Figure~\ref{fig:poly} (a).
The collapse \textit{looks} visually continuous, although the images show a collapse mediated by multiple intermediate, discrete steps, and the shape of the order parameter with time is sigmoidal.
Figure~\ref{fig:poly} (c - left) shows the cost associated with the choice of optimum $n$ (number of change points) for offline detection.
The offline detection scheme only detects the transition to the final snapshot of the simulation because fitting to the individual sections of the collapse does not sufficiently decrease the cost function.

Figure~\ref{fig:poly} (c - right)  shows \pkg{dupin}'s ability to detect the collapse and its component parts in the case of online detection.
For online detection the elbow cost plot would need to be shown for every frame.
Instead, we show the relative improvement of cost inside the current window from adding a second change point in online detection.
The formula used is $\xi = (c_1 - c_2) / c_0$ where  $c_i$ is the cost for selecting $i$ change points.
This approach yields independent cost curves for each window (detected event).
When this proxy is high, it indicates that the current window is likely to contain a transition.
The ``ramp up'', ``growth'' and ``slow down'' behavior of the sigmoid leads to three events detected with online detection at frames 65, 95, and 140.
This granularity results from the max signal length of 50, which increases the relative reduction in cost for fitting to the three sections of the polymer collapse.
Processing the entire trajectory for online detection took 820 milliseconds ($\pm 3.12 \mu s$) on a single core of a 3.0 GHz Intel Xeon Gold 6154.
This speed is sufficiently fast to use in real time applications such as autonomously triggering simulation protocols during a simulation.

\section{Discussion and Conclusions}%
\label{sec:discussion}
We've demonstrated how the procedure described in Section~\ref{sec:scheme} allows for detection of transition points within a simulation trajectory with a high degree of accuracy.
The obvious benefit from this approach is the automation of structural transition detection within a study.
In studies with hundreds and thousands of simulations the dividends of this approach increase exponentially.
We have also demonstrated that \pkg{dupin} is successful at detecting events underwent by a small number of particles in the system ($<5\%$).
Our method does, however, require informative descriptors for the transition.
This requirement can be met by selecting a wide variety of descriptors and applying feature selection tools to the signal afterwards.

From the examples presented, we conclude that the ML dimensionality reduction classifier produces a signal which results in a larger partitioning of the detected events compared to results produced using the linear cost function approach.
However, the ML dimensionality reduction classifier approach falls short in cases where multiple events are overlapping or in cases where there are abrupt changes at the end of the signal.
For instance, if we were to apply the ML dimensionality reduction classifier approach to a system with only one frame into a transition, no event would be detected.
The linear approach is better suited for detection in such scenarios and in scenarios where multiple events are expected as showcased in the online polymer example.

\pkg{dupin} opens the doors for new ML applications to phenomena such as crystallization pathways, defect formation and active matter, all of which involve structural transitions.
By curating the transition data for ML applications, large scale studies that would have been prohibitively costly in terms of human hours are now accessible.
Furthermore, leveraging \pkg{dupin} for online event detection holds promise to lessen data storage and processing demands and provide a powerful avenue for real-time control over simulations and experiments.

The source code can be found at GitHub at \url{https://github.com/glotzerlab/dupin}.
The documentation hosted by Read the Docs, \url{https://dupin.readthedocs.io}, also contains three additional examples of event detection: one for a simple example of WCA spheres forming FCC, one of hard truncated tetrahedra forming cF432, and one of ionic beryllium and chlorine forming a nematic then crystalline phase.
The analysis code and data will be available on Deep Blue Documents following publication.
\section{Acknowledgements}%
\label{sec:ack}
This research was supported by the National Science Foundation, Division of Materials Research under a Computational and Data-Enabled Science and Engineering (CDS\&E) Award \# DMR 2302470 (2023 - 2027) and \# DMR 1808342 (2019-2023).
This work used the Extreme Science and Engineering Discovery Environment (XSEDE)~\cite{towns.etal2014}, which is supported by National Science Foundation grant number ACI-1548562; XSEDE award DMR 140129.
This research was supported in part through computational resources and services provided by Advanced Research Computing, a division of Information and Technology Services at the University of Michigan, Ann Arbor.

    \printbibliography
    \makeatletter\@input{si-aux.tex}\makeatother
\end{document}